\documentclass[10pt,journal,twoside]{IEEEtran}

\usepackage{graphicx}
\usepackage{color}
\usepackage{placeins}
\usepackage{float}
\usepackage{tabularx,colortbl}
\usepackage{multirow}
\usepackage{makecell}
\usepackage{CJK}
\usepackage{amsmath,bm}
\usepackage{cite}
\usepackage{amsthm}
\usepackage{extarrows}
\usepackage{amssymb}
\usepackage[table]{xcolor}

\usepackage[ruled,vlined]{algorithm2e}
\usepackage{placeins}
\usepackage{tabularx,colortbl}
\usepackage{multirow}
\usepackage{makecell}
\usepackage{CJK}
\usepackage{bm}
\usepackage{amsthm}
\usepackage{extarrows}
\usepackage{amssymb}
\usepackage[table]{xcolor}

\usepackage{amsmath}

\begin{document}
%
\title{A New Image Codec Paradigm for Human and Machine Uses}
%
%

\author{Sien Chen, Jian Jin,~\IEEEmembership{Member,~IEEE}, Lili Meng, Weisi Lin,~\IEEEmembership{Fellow,~IEEE}, 

Zhuo Chen, Tsui-Shan Chang, Zhengguang Li, Huaxiang Zhang

\thanks{Copyright \copyright 20XX IEEE. Personal use of this material is permitted. However, permission to use this material for any other purposes must be obtained from the IEEE by sending an email to pubs-permissions@ieee.org. \emph{(Corresponding author: Jian Jin, LiLi Meng.)}}
\thanks{S. Chen, L. Meng, and H. Zhang are with the School of Information Science and Engineering, Shandong Normal University, Jinan, 250014, China. E-mail: 2019020854@stu.sdnu.edu.cn; mengll\_83@hotmail.com; huaxzhang@hotmail.com.}
\thanks{J. Jin, W. Lin, Z. Chen are with the School of Computer Science and Engineering, Nanyang Technological University, 639798, Singapore and also with Alibaba-NTU Singapore Joint Research Institute, Nanyang Technological University, 639798, Singapore. E-mail: jian.jin@ntu.edu.sg; wslin@ntu.edu.sg; zchen036@e.ntu.edu.sg.}
\thanks{T.-S. Chang and Z. Li are with the Alibaba cloud business group, Alibaba, Hangzhou 310052, China. Email: tsuishan.cts@alibaba-inc.com; zhengguang.lzg@alibaba-inc.com.}}

\markboth{IEEE Transactions on Circuits and Systems for Video Technology.}%
{Sien \MakeLowercase{\textit{et al.}}:
~A New Image Codec Paradigm for Human and Machine Uses}
%



\maketitle
\begin{abstract}
With the AI of Things (AIoT) development, a huge amount of visual data, e.g., images and videos, are produced in our daily work and life. These visual data are not only used for human viewing or understanding but also for machine analysis or decision-making, e.g., intelligent surveillance, automated vehicles, and many other smart city applications. To this end, a new image codec paradigm for both human and machine uses is proposed in this work. Firstly, the high-level instance segmentation map and the low-level signal features are extracted with neural networks. Then, the instance segmentation map is further represented as a profile with the proposed 16-bit gray-scale representation. After that, both 16-bit gray-scale profile and signal features are encoded with a lossless codec. Meanwhile, an image predictor is designed and trained to achieve the general-quality image reconstruction with the 16-bit gray-scale profile and signal features. Finally, the residual map between the original image and the predicted one is compressed with a lossy codec, used for high-quality image reconstruction. With such designs, on the one hand, we can achieve scalable image compression to meet the requirements of different human consumption; on the other hand, we can directly achieve several machine vision tasks at the decoder side with the decoded 16-bit gray-scale profile, e.g., object classification, detection, and segmentation. Experimental results show that the proposed codec achieves comparable results as most learning-based codecs and outperforms the traditional codecs (e.g., BPG and JPEG2000) in terms of PSNR and MS-SSIM for image reconstruction. At the same time, it outperforms the existing codecs in terms of the mAP for object detection and segmentation.   


\end{abstract}

\IEEEpeerreviewmaketitle

\section{Introduction}
\IEEEPARstart{I}{mage} compression is the main component in signal processing. Its related technologies have been widely studied and applied for various applications for decades. The goal of image compression technologies is to achieve an efficient representation of the image during image communication and storage. In the early researches, image compression technologies (e.g., JPEG\cite{wallace1992the}, JPEG2000\cite{taubman2002jpeg2000}, BPG\cite{bellard2015bpg}, FLIF\cite{sneyers2016flif}, VVC\cite{vvc} and\cite{toderici2016variable,balle2017end,toderici2017full,johnston2018improved,theis2017lossy,agustsson2017soft,balle2018variational,minnen2018joint,cheng2019Deep} etc.) were designed for human viewing purposes (human consumption), since humans were the final receptors of the decoded image for most applications, such as digital photography, social media, etc. At that time, images were compressed to achieve a good balance between low bit rate cost and high-quality visual signal reconstruction. Afterward, people developed some more efficient ways to compress the image with visual descriptors technologies (e.g., SIFT\cite{lowe2007distinctive}, CDVS\cite{Djamasbi2004compact}) for certain machines tasks (machine consumption). Their designs were machine vision tasks specified, e.g., specify visual feature extraction and compression for certain image analysis tasks, which made visual analysis tasks performed more efficiently. The goal of image compression became to achieve a good balance between low bit rate cost and high-performance of machine vision tasks. However, such kind of representation can not be directly understood by humans. 

With the rise of the Internet of Things (IoT) and deep learning, more and more edge devices are equipped in the smart city to enable higher-level automation, generating a huge amount of visual data (images and videos) to be transmitted and processed in real-time. Visual data are consumed by both humans (e.g., human viewing) and machines (e.g., visual analysis in autopilot, defect detection and quality assurance for advanced manufacturing, abnormal detection for smart healthcare, etc). In view of this, image compression should simultaneously address the needs of human and machine consumption. However, the aforementioned image compression techniques can not fulfill these two kinds of requirements both. For instance, traditional image compression technologies are designed to ensure high-quality reconstructed visual signals appealing to human visual perceptual but not optimized for machine vision tasks. Visual descriptors technologies were designed for specified machine vision tasks. The conveyed data is abstract and unreadable for human. 

To address the drawbacks above, some related works have been proposed \cite{BO2019direct,qian2012an,khatoonabadi2013video,choi2017corner,choi2017hevc,ranjbar2018can,torfason2018towards,alvar2019multi,alvar2020bit,hu2020towards,liu2021semantics,choi2021latent,patwa2020semantic}. Commonly, the bitstream of the entire image is used to perform machine vision tasks, such as face detection \cite{ ranjbar2018can}, human detection \cite{choi2017hevc} and so on, while the decoded image is used for human viewing. Although machine vision tasks can be directly performed at the bitstream level and the decoding process is saved in this situation, the bitstream of the entire image is still needed for machine vision tasks. Such kind of design cannot achieve scalable image compression. Considering this, several works \cite{hu2020towards,liu2021semantics,choi2021latent,yan2021scalable} try to represent the entire image with multiple bitstreams. Part of the bitstream is decoded and used for machine analysis tasks, while the image can be fully decoded for human consumption when all the bitstreams are obtained. However, such kinds of codecs can not achieve good performance when both human and machine consumption are considered, especially for machine vision tasks.


In this paper, a new image codec paradigm is proposed for both human and machine uses, which is a predictive coding framework inspired by \cite{akbari2019dsslic}. To this end, the high-level instance segmentation map and low-level signal features are extracted at first. Unlike directly compressing these two kinds of information into multiple bitstreams with lossy image codecs in \cite{hu2020towards,choi2021latent,yan2021scalable}, the instance segmentation map is represented as a 16-bit gray-scale profile. Then, it together with the signal features are compressed with a lossless codec. To achieve the general-quality image prediction/reconstruction with the 16-bit gray-scale profile and signal features, we design an image predictor. After that, the residual map between the original image and its associated predicted one (generated via the image predictor) is compressed with a lossy codec to achieve high-quality image reconstruction. All these designs make the proposed method achieve scalable image compression. The main contributions are summarized as follows:

\begin{itemize}

\item We extract the instance segmentation map instead of the features or other information used in \cite{akbari2019dsslic, choi2021latent} as the high-level semantic information. It allow us to achieve object detection and segmentation directly. Besides, the valuable semantic information (spatial location, edge information, etc.) existed in the instance segmentation map are also used to help with image reconstruction. 

\item We propose a 16-bit gray-scale representation technique, which can represent the instance segmentation map as a more compact 16-bit gray-scale profile. Besides, the proposed technique also takes the category of object into account. With 16-bit gray-scale, we can directly achieve object classification, detection and segmentation at the decoder side.

\item We use a combination of lossless and lossy compression. The lossless compression applied on the 16-bit gray-scale profile guarantee almost no performance degradation of three machine vision tasks, especially for low bit rate situations, which is critical for lots of machine vision applications, such as autopilot, smart robot, etc. The lossless compression of low-level signal features also preserves the significant information for general-quality image reconstruction. By applying lossy compression on the residual map, we get the PSNR-BPP, MS-SSIM-BPP, and mAP-BPP curves of the entire bitstreams to satisfy different bandwidth. 


\item Considering that normalization plays a critical role in data representation in the proposed extractors and predictor, we widely investigate the commonly used normalization methods. By integrating the channel-wise normalization into our networks, the compression efficiency is further improved, which also demonstrate that channel normalization is more efficient for the predictive coding paradigm. 
\end{itemize}


By applying all these technologies above, we achieve comparable results as most deep learnt codecs and outperform the traditional codecs (e.g., BPG and JPEG2000) for image reconstruction. While, it outperforms the existing codecs for machine vision tasks, especially for the low bit rate situations. 

The outline of the rest of our paper is as follows. Section II reviews the related works on image codec. Section III proposes an image codec paradigm for human and machine uses. Section IV presents experimental results and Section V concludes this paper.

\section{Related work}

In this section, we give a review of the development of image codecs along with three directions, i.e., image compression for human consumption, machine consumption, and both human and machine consumption. 

\subsection{Image compression for humans}
At the early research of image compression, image compression aimed to achieve high-quality signal reconstruction by the constraint of bit-rate, namely serving for human consumption. The traditional hybrid image compression paradigm (such as the paradigm used in JEPG\cite{wallace1992the} and JPEG 2000\cite{taubman2002jpeg2000}) mainly contained three steps, i.e., transform, quantization, and entropy coding. Such kind of design guaranteed high performance in signal fidelity-inspired metrics (e.g., PSNR, SSIM\cite{wang2003multi}, etc.) and also achieved good results in perceptual metrics (e.g., JND \cite{JND}) to some degree. 

With the rising of deep learning, lots of works\cite{toderici2016variable,balle2017end,toderici2017full,johnston2018improved,theis2017lossy,agustsson2017soft,balle2018variational,minnen2018joint,cheng2019Deep} started to address the image compression problem with the learning ways. However, they still aimed to optimize their neural networks towards high-quality signal reconstruction. For instance, they try to use the CNN technologies to replace the DCT transform used in the traditional hybrid image codec to achieve a compact representation of the image. 
Meanwhile, Some generative adversarial models \cite{rippel2017real,Santurkar2018g,agustsson2019generative} were utilized to learn the distribution of images to obtain better subjective visual quality at extremely low bit rates. Besides, importance maps was used as an alternative to discrete entropy estimation to control the compression ratio in \cite{li2018l,chneg2018d,cheng2019Deep}. 
After that, context models \cite{balle2018variational,minnen2018joint,lee2017context,mentzer2018c} for entropy estimation were developed, which are used for guiding the optimization of neural network parameters, balancing the trade-off between signal distortion and required bits (entropy). Subsequently, a deep semantic segmentation-based layered (DSSLIC) image compression algorithm was proposed in \cite{akbari2019dsslic}, where the semantic information from the input image is first extracted as the basic layer to achieve low quality image reconstruction. Then, the residual information together with side information are compressed as an enhanced layer. It was the first work to involve semantic information for image reconstruction and achieved scalable image compression with the predictive coding framework. The learning-based methods above greatly improved the efficiency of image compression. In some cases, they even surpass the industrial hybrid codecs, such as BPG (HEVC single frame codec). 

\begin{figure*}[htbp]
	\begin{center}
		\noindent
		\includegraphics[width = 6.5 in]{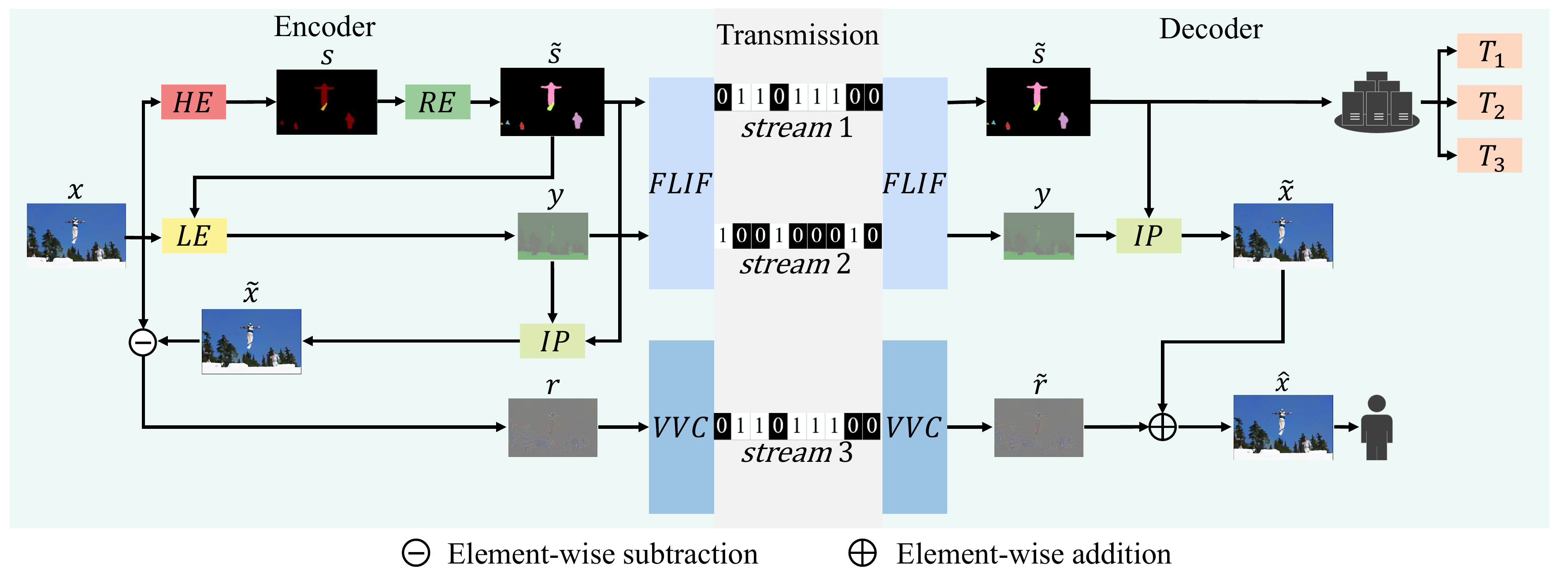}
		\caption{The framework of a new paradigm on learning image codecs for human and machine uses, which is mainly composed of two parts: encoder and decoder. At the encoder side, the original image $x$ is first fed into \emph{HE} (Mask-RCNN) to extract the instance segmentation map as the high-level semantic feature $s$. Subsequently, $s$ is represented as a 16-bit gray-scale profile $\tilde{s}$ via the the proposed module \emph{RE}. Then, $x$ and $\tilde{s}$ are fed into \emph{LE} to extract low-level signal information $y$. Next, $\tilde{s}$ and $y$ are sent to the module \emph{IP} to obtain the general-quality image $\tilde{x}$. After that, the residual map $r$ between $x$ and $\tilde{x}$ is generated by applying element-wise subtraction operation on $x$ and $\tilde{x}$. Then, $\tilde{s}$ and $y$ are losslessly compressed via \emph{FLIF} and obtain two bitstreams, i.e., $stream1$ and $stream2$. Meanwhile, $r$ is lossy compressed via \emph{VVC} to generate the third bitstream $stream3$. After transmission through channel, the 16-bit gray-scale profile $\tilde{s}$, low-level signal feature $y$, and residual map $r$ are firstly decoded from $stream1$, $stream2$, and $stream3$ at the decoder side. On the one hand, three machine vision tasks can be achieved by obtaining the object category, bounding box, and instance segmentation mask from $\tilde{s}$, such as object classification $T_1$, object detection $T_2$, and instance segmentation $T_3$. On the other hand, the decoded $y$ and $\tilde{s}$ can be used to generate the general-quality image $\tilde{x}$ via the module \emph{IP}. After integrating the decoded $\tilde{r}$ with $\tilde{x}$ via the element-wise addition operation, a high-quality image $\hat{x}$ can be reconstructed.}\label{fig_FW}
	\end{center}
\end{figure*}

However, both traditional hybrid codecs and the rising learning-based codecs cannot achieve good performance for machine vision tasks, as they are designed for image reconstruction by nature. 

\subsection{Image compression for machines}

Description coding, also know as image feature coding, was proposed to generate compact image descriptions (e.g., SIFT\cite{lowe2007distinctive} and CDVS\cite{Djamasbi2004compact}) for efficient data transmission, machine analysis and storage. It is desirable for machine vision in case of low bit rate, as image signals usually suffer from compression artifacts by conventional image coding which results in poor analysis task performance \cite{dodge2016u}. Moreover, different from conventional image coding, machine-orientated image compression only conveys necessary information for subsequent machine consumption, which is less expensive and of higher secure level. For example, a learning-based semantic structure coding (SSC) was proposed in \cite{he2019beyond,sun2021semantic}, where an object detection module in the feature domain is used to locate and align objects and then reorganize the features with such information for generate a semantically structured bitstream (SSB). As a result, it can directly obtain the information required by a variety of machine analysis tasks from the designated area in the SSB, instead of transmitting a complete bitstream. Recently, the transferability between different tasks was used in \cite{hy2021revisit} to construct a more efficient representation, combined with a codebook-based hyperprior, this model can support a variety of machine analysis tasks at a very low bpp. On the other hand, the standardization organization MPEG also announced a new generation of compression descriptors MPEG-CDVA \cite{sun2020cdva} for video analysis. However, the aforementioned Description coding methods can not work for image reconstruction tasks.

Compared with the image codecs optimized for humans, image description coding for machines can support a variety of machine tasks with small bitstream. However, in case of applications including both machine analysis and human viewing, another type of coding scheme should be explored.

\subsection{Image compression for human and machine uses}


A straightforward way to deal with both human viewing and machine analysis is to naively compress the image signals and take the reconstructed image for machine and human consumption. However, this simple implementation may brings many problems. For examples, performance of machine analysis tasks will be significantly reduced at low bit rate; original image signal usually contains too much redundant information for machine analysis. Therefore, how to design a compression pipeline that can reduce data volume without significantly degrading quality of reconstructed image and task performance for machine analysis has become a problem.

As to the compression pipeline design, some early works \cite{BO2019direct,qian2012an,choi2017corner} tried to perform machine analysis tasks in the DCT domain, undecoded and partially decoded HEVC bitstreams. 
These tasks include edge detection, human recognition, and generation of corner proposals that can be used for object recognition, tracking, and retrieval. In addition, a convolutional neural network is used to perform face detection on the entropy-decoded HEVC bitstream in \cite{ranjbar2018can}. Subsequently, machine analysis tasks such as classification and segmentation were directly performed in the compressed representation generated by the autoencoder in \cite{torfason2018towards}. The aforementioned methods bypass the process of mapping the entropy-decoded bitstream to the RGB domain but still need to transmit the entire image bitstream.


To reduce the bitstream transmitted for machine tasks, a multi-task collaborative intelligence latent space scalability was proposed in \cite{choi2021latent}. The authors use the decoding part of the latent space to support object detection, and the remaining space is used for input reconstruction. Subsequently, a three-layer model based on the backbone of Feature Pyramid Network (FPN) was designed in \cite{choi2021scalable}, which extended the supported tasks to object detection, object segmentation, and input reconstruction. 
Recently, a SSSIC scheme was proposed in \cite{yan2021scalable}, integrating the semantics and signal representation into one single framework. The author using a coarse-to-fine method to extend the supported machine analysis tasks to object classification, object detection, and instance segmentation. However, all the works above are the initial exploration on image compression for human and machine uses. More deeper researches are still needed.



\section{A New Image Codec Paradigm for Human and Machine Uses}

In this section, we first introduce the framework of the proposed image codec paradigm for human and machine uses. Then, the relevant technologies used in the main components of the paradigm are elaborated in turns, which includes the architectures of the proposed feature extractors and image predictor as well as the proposed 16-bit gray-scale representation.  

\subsection{Framework}

At the encoder side, the original image, denoted by $x$, is the input. Firstly, $x$ is fed into \emph{HE} to extract the instance segmentation map as the high-level semantic features, denoted by $s$. This process is represented by 
\begin{equation}
\label{MaskRCNN}
s = H(x; \xi_H),  
\end{equation}
where $H(\cdot; \cdot)$ represents the pre-trained Mask-RCNN networks and $\xi_H$ represents the pre-trained fixed parameters. 

Then, $s$ is represented as a 16-bit gray-scale profile $\tilde{s}$ via the proposed \emph{RE} module and we have
\begin{equation}
\label{RE1}
\tilde{s} = R(s),  
\end{equation} 
where $R(\cdot)$ denotes the 16-bit gray-scale representation.  

Then, $x$ together with $\tilde{s}$ are fed into the \emph{LE} to extract the low-level signal features, denoted by $y$, which can be formulated as follow,
\begin{equation}
\label{LE}
y=L(\tilde{s}, x; \xi_L), 
\end{equation} 
where $L(\cdot, \cdot; \cdot)$ denotes the \emph{LE} network and $\xi_L$ represents its trainable parameters.

After that, $\tilde{s}$ is used as the side information. After combined with $y$, a general-quality image $\tilde{x}$ is obtained via the image predictor \emph{IP}, denoted by $I(\cdot, \cdot; \cdot)$. This process is represented by
\begin{equation}
\label{IP}
\tilde{x}=I(\tilde{s},y;\xi_I),
\end{equation} 
where $\xi_I$ denotes trainable parameters of $I(\cdot, \cdot; \cdot)$. 

Then, the residual map $r$ between original image $x$ and its predicted one $\tilde{x}$ is generated via the element-wise subtraction operation \textcircled{-}. Both $\tilde{s}$ and $y$ are losslessly compressed with \emph{FLIF} and generate two bitstreams, denoted by $stream1$ and $stream2$. Meanwhile, $r$ is lossy compressed with \emph{VVC} and generate the third bitstream, denoted by $stream3$. Finally, these three bitstreams are transmitted to the decoder via the channel. It should be noticed that $\tilde{s}$ is a compact representation of $s$, which contains the object category, hardly degraded bounding box and instance segmentation mask. The details of 16-bit gray-scale representation will be elaborated in subsection \ref{RE}.

At the decoder side, 16-bit gray-scale profile $\tilde{s}$, low-level signal features $y$, and residual map $r$ are firstly decoded from the $stream1$, $stream2$ and $stream3$, respectively. On the one hand, three major machine vision tasks, i.e., object classification, object detection, and instance segmentation tasks, can be achieved by obtaining the object category, bounding box, and instance segmentation mask from $\tilde{s}$. On the other hand, the decoded $y$ together with $\tilde{s}$ are used to reconstruct the general-quality image $\tilde{x}$ via the module \emph{IP}. This process can be represented as 
\begin{equation}
\label{IP_De}
\tilde{x}=I(F(\tilde{s}),F(y);\xi_I),
\end{equation} 
where $F(\cdot)$ denotes the encoding-decoding process via codec \emph{FLIF}. As it's a lossless compression process, Eq. \eqref{IP_De} can be replace be Eq. \eqref{IP} here.   

However, $r$ is compressed via the \emph{VVC}, which is lossy compression process. We have
\begin{equation}
\label{vvc}
\tilde{r}=V(r),
\end{equation} 
where $V(\cdot)$ denotes the lossy encoding-decoding process via codec \emph{VVC}. 

After integrating $\tilde{r}$ with $\tilde{x}$ via an element-wise addition operation \textcircled{+}, a high-quality image $\hat{x}$ is obtained, which can be formulated as,
\begin{equation}
\label{high}
\hat{x}=\tilde{x} + \tilde{r}.
\end{equation} 
It should be noticed that either general-quality image $\tilde{x}$ or high-quality image $\hat{x}$ is exhibited for human consumption. It depends on the bandwidth and the requirements of the customers. In this work, considering that the \emph{VVC} compression process is nondifferentiable, we select $\tilde{x}$ as our target to constrain instead of $\hat{x}$. For $\tilde{x}$, we have
\begin{figure}[htbp]
	\begin{center}
		\noindent
		\includegraphics[width = 2.8 in]{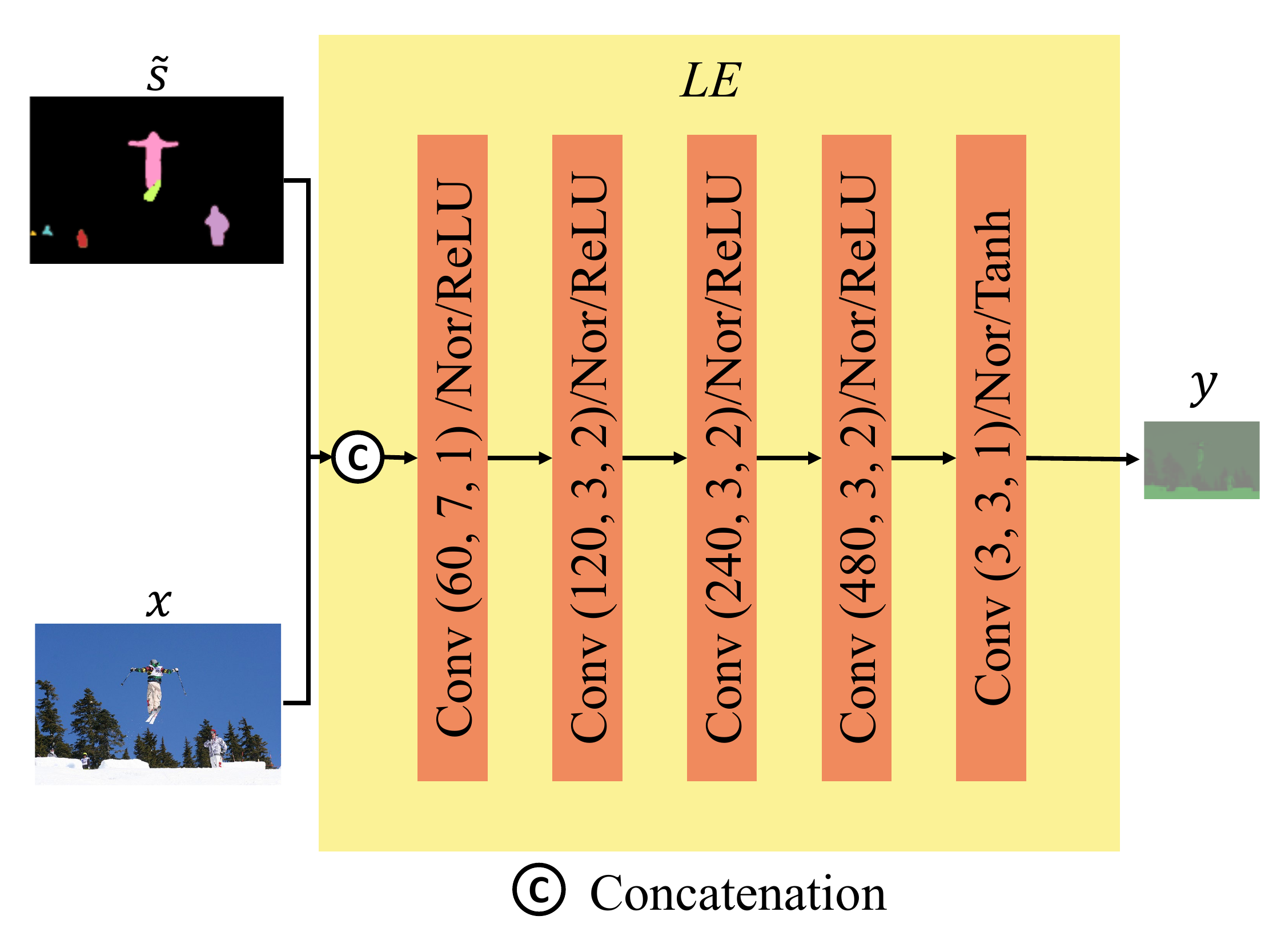}
		\caption{The architecture of low-level signal extractor (\emph{LE}).}
		\label{fig_LE}
	\end{center}
\end{figure}
\begin{equation}
\label{long}
\begin{split}
\tilde{x} &= I(F(\tilde{s}), F(y); \xi_I)\\
&= I(\tilde{s}, y; \xi_I)\\
&= I(\tilde{s}, L(x,\tilde{s};\xi_L); \xi_I)\\
&= I(R(H(x;\xi_H)), L(x,R(H(x;\xi_H));\xi_L); \xi_I). \\
\end{split}
\end{equation}
Since the parameter $\xi_H$ of Mask-RCNN is a fixed value, the performance of the proposed model is dependent on $\xi_L$ and $\xi_I$. During $L(\cdot,\cdot;\cdot)$ and $I(\cdot,\cdot;\cdot)$ optimization, we use $l_1$ and $SSIM$ \cite{wang2004image} loss to constrain to guarantee that $\tilde{x}$ can approach to $x$ as close as possible. For $l_1$ loss, we have 
\begin{equation}
\label{long1}
\mathcal{L}_{1} = \lambda \cdot \|x - \tilde{x} \|_ {1}.
\end{equation}
For $SSIM$ loss, we have 
\begin{equation}
\label{ssim}
\mathcal{L}_{2}=-\beta \cdot \frac{2 \mu_{x} \mu_{\tilde{x}}+C_{1}}{\mu_{x}^{2}+\mu_{\tilde{x}}^{2}+C_{1}} \cdot \frac{2 \sigma_{x \tilde{x}}+C_{2}}{\sigma_{x}^{2}+\sigma_{\tilde{x}}^{2}+C_{2}},
\end{equation}

where $C_{1}=(255 \cdot K_{1})^{2}$, $C_{2}=(255\cdot K_{2} )^{2}$. Here, we set $K_1$ and $K_2$ to 0.01 and 0.03, respectively. $\mu_ {x}$ represents the mean value of pixel values in $x$. $\sigma_{x \tilde{x}}$ represents the covariance between $x$ and $\tilde{x}$. $\sigma_{x}$ is the variance of $x$. 

Then, we have the objective function as follows,
\begin{equation}
\label{loss}
\begin{split}
\mathcal{L}=\lambda \cdot \|x-\tilde{x}\|_{1} - \beta \cdot \frac{2 \mu_{x} \mu_{\tilde{x}}+C_{1}}{\mu_{x}^{2}+\mu_{\tilde{x}}^{2}+C_{1}} \cdot \frac{2 \sigma_{x \tilde{x}}+C_{2}}{\sigma_{x}^{2}+\sigma_{\tilde{x}}^{2}+C_{2}},
\end{split}
\end{equation}

where $\lambda$ and $\beta$ are set to $2$ and $1$, respectively.

\subsection{Architectures of module LE and IP}
\label{LE_IP}

\subsubsection{Architecture of module LE}
The architecture of low-level signal extractor (\emph{LE}) is shown in Fig. \ref{fig_LE}, which contains three parts, namely the inputs, the convolutional layers, and the outputs. As aforementioned, the inputs are the profile of the high-level instance segmentation map $\tilde{s}$ and original image $x$, respectively. Firstly, $\tilde{s}$ and $x$ are stacked along channel dimension with concatenation operation, denoted by \textcircled{c}. Then, they are followed by five convolutional layers $Conv(A,B,C)$, normalization $Nor$, and activation function $ReLU/Tanh$. Notations $A$, $B$ and $C$ are the number of kernels, the size of filter, 
\begin{figure}[htbp]
	\begin{center}
		\noindent
		\includegraphics[width = 3.5in]{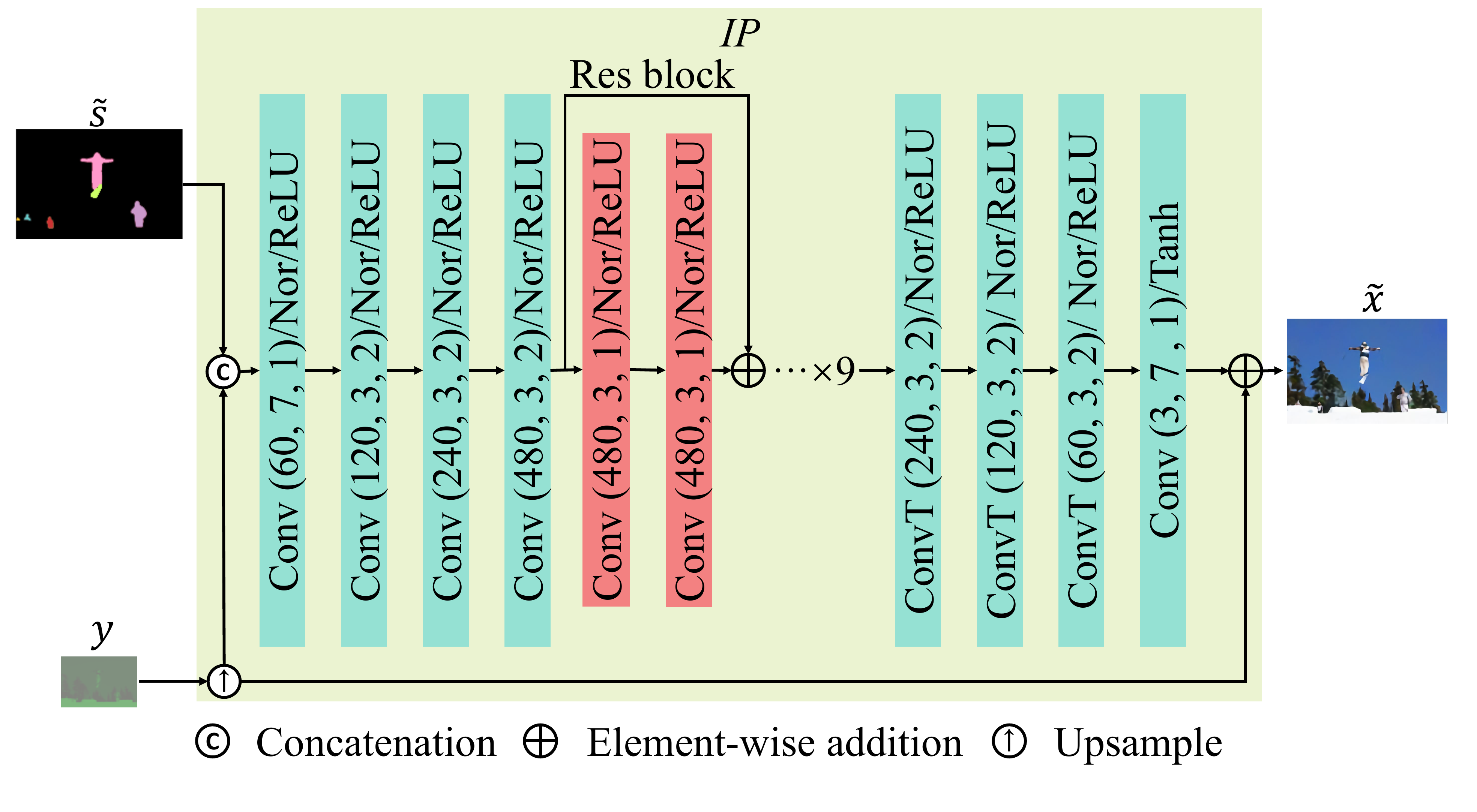}
		\caption{The architecture of original image predictor (\emph{IP}).}\label{fig_IP}
	\end{center}
\end{figure}and the stride of the slide, respectively. At the first layer, a larger filter (e.g., $B$ = $7\times7$) with small stride (e.g., $c$ = 1) is used to provide large enough receptive field and helps to extract more global features of the inputs. For the followed three layers, a small filter with larger stride is used to compress the spacial redundancy of the features, while the number of filters are enlarged for the first four layers to enrich the features. At the last layer, the redundancy of the features along channel dimension is largely compressed to obtain a more compact representation of the low-level signal information, where only $3$ main channels of the low-level signal features are preserved as the outputs $y$.  

\subsubsection{Architecture of module IP}
As shown in Fig. \ref{fig_IP}, the main architecture of original Image Predictor (\emph{IP}) is a convolutional auto-encoder like design, which is mainly made up convolutional encoder and transpose-convolutional decoder.

Firstly, the low-level signal features $y$ is upsampled (nearest) to the size of original image via the unsample operation. Then, the upsampled map together with $\tilde{s}$ are stacked via the concatenation operation. Then, they are filtered by four convolutional layers, nine residual blocks, and four transpose convolutional layers. Similarly, each of them followed by normalization $Nor$ and activation function $ReLU/Tanh$. Finally, the filtered features are added to the upsampled map via element-wise addition operation and the general-quality image $\tilde{x}$ is obtained. This kind of design aims to well predict the residual information between upsampled map of $y$ and original image $x$ to recover the over-compressed information at the \emph{LE}.

In this work, only $3$ main channels of the low-level signal features are preserved at the \emph{LE} for the following image reconstruction at the \emph{IP}, which is much less than that of the bottleneck features used in the existing end-to-end image codecs. It means much more redundancy along channel dimension (maybe include some useful information for image reconstruction) is removed at the \emph{LE} and the over-compressed information may be recovered at the \emph{IP} during image reconstruction. In view of this, we apply channel-wise normalization instead of batch-/instance-wise normalization at each layers of \begin{figure}[htbp]
	\begin{center}
		\noindent
		\includegraphics[width = 3.0in]{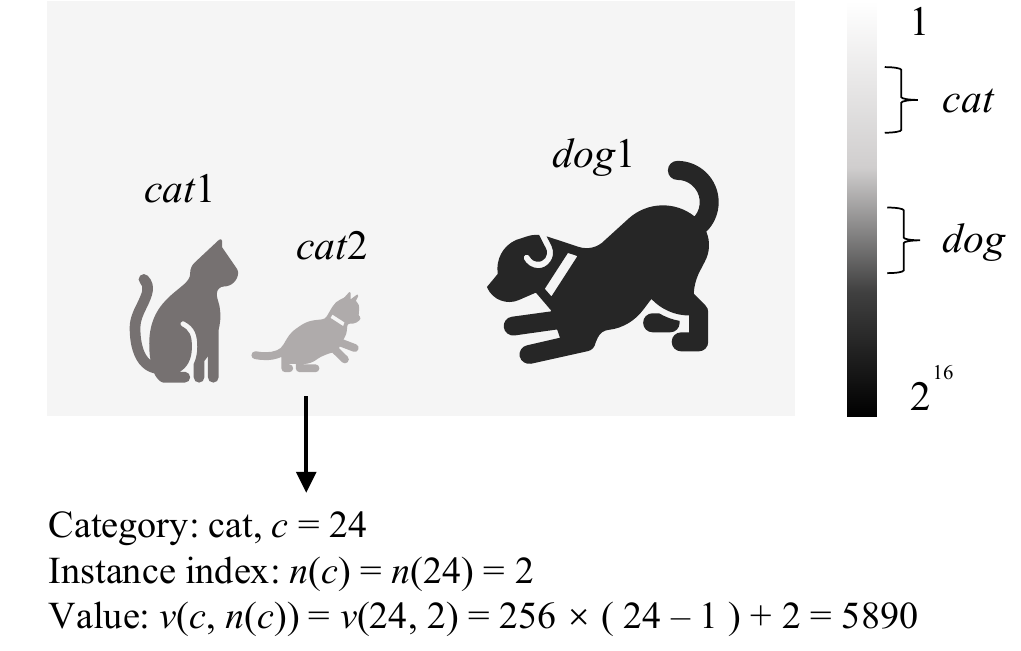}
		\caption{The proposed 16-bit gray-scale representation (\emph{RE}).}\label{fig_RE}
	\end{center}
\end{figure}
\emph{LE} and \emph{IP}. Experimental results in subsection \ref{nor} also demonstrate that channel-wise normalization brings additional improvement of the compression efficiency of the proposed image codec paradigm.

\subsection{Design of module RE}
\label{RE}


Commonly, the instance segmentation map is an RGB colorful image. Different colors are used to distinguish different instances (objects). In extremes, there are $256$ values for each color channel. For RGB three channels, $256 \times 256 \times 256$ colors ($24$ bits) are used to distinguish different instances, which is not friendly for compression. Besides, the category and its correspondingly colors of masks are recorded in a dictionary. Different datasets may use different dictionaries to represent the relation between category and its correspondingly colors of masks. Therefore, even though we get the instance image at the decoder side, we cannot get the category of each masked instance without the unique dictionary. 

In view of this, a 16-bit gray-scale representation \emph{RE} is proposed here. Before that, we investigate the number of the categories and instances in different dataset. E.g., for COCO dataset, there are $80$ categories. In the extreme case, there are $96$ instances with the same category label in an image. The other dataset are similar. Therefore, we use $8$ bits to represent the categories of objects existed in an image and obtain an uniform dictionary. For each category, we use $8$ bits to distinguish different objects. Hence, the mask of any instance can be represented with 16-bit gray-scale value. Assume there are $256$ instances with the same category $c$. For the $n$-th instance with category $c$, its gray-scale value $v$ can be represented as follows,

\begin{equation}
\label{mapping}
v(c,n(c)) = 256 \cdot (c - 1) + n(c).  
\end{equation}

To make it easier to understand, we also give an example on 16-bit gray-scale profile as shown in Fig. \ref{fig_RE}. There are two cats and one dog in the profile. According to our uniform dictionary, assume category $cat$ and $dog$ are represented with number $24$ and $45$, respectively. Then, the gray-scale value of $cat1$ is $5889 = 256 \times (24-1) + 1$. Similarly, the gray-scale values of $cat2$ and $dog1$ are $5890 = 256 \times (24-1) + 2$, $11265 = 256 \times (45-1) + 1$, respectively. Therefore, the 24 bits RGB instance segmentation map can be represented with a 16-bit gray-scale profile, which is more compressible for the followed lossless compression. 

At the decoder side, the category of instances can be obtained by applying the following equation,
\begin{equation}
\label{invers-mapping}
c = \lfloor \frac{v(c,n(c))}{256} \rfloor + 1,  
\end{equation}
where $\lfloor \cdot \rfloor$ is the flooring calculation. Therefore, for the $cat2$ in Fig. \ref{fig_RE}, $c = \lfloor \frac{5890}{256} \rfloor + 1 = 24$. Similarly, the categories of the $cat1$ and the $dog1$ are $24$ and $45$. By looking up our uniform dictionary, $24$ and $45$ represent category $cat$ and $dog$, respectively. Therefore, with such profile, the category of each instance can be fully recovered. Considering the efficiency of the compression, the floating coordinates of each instance are rounded to the integer during the representation. However, this leads to insignificant degradation when the instance segmentation map is reconstructed. All these designs allow us to achieve object classification without performance degradation while achieving object detection and instance segmentation with insignificant performance degradation.

\section{Experiments}

%
\begin{figure*}[htbp]
	\begin{center}
		\noindent
		\includegraphics[width = 7.1in]{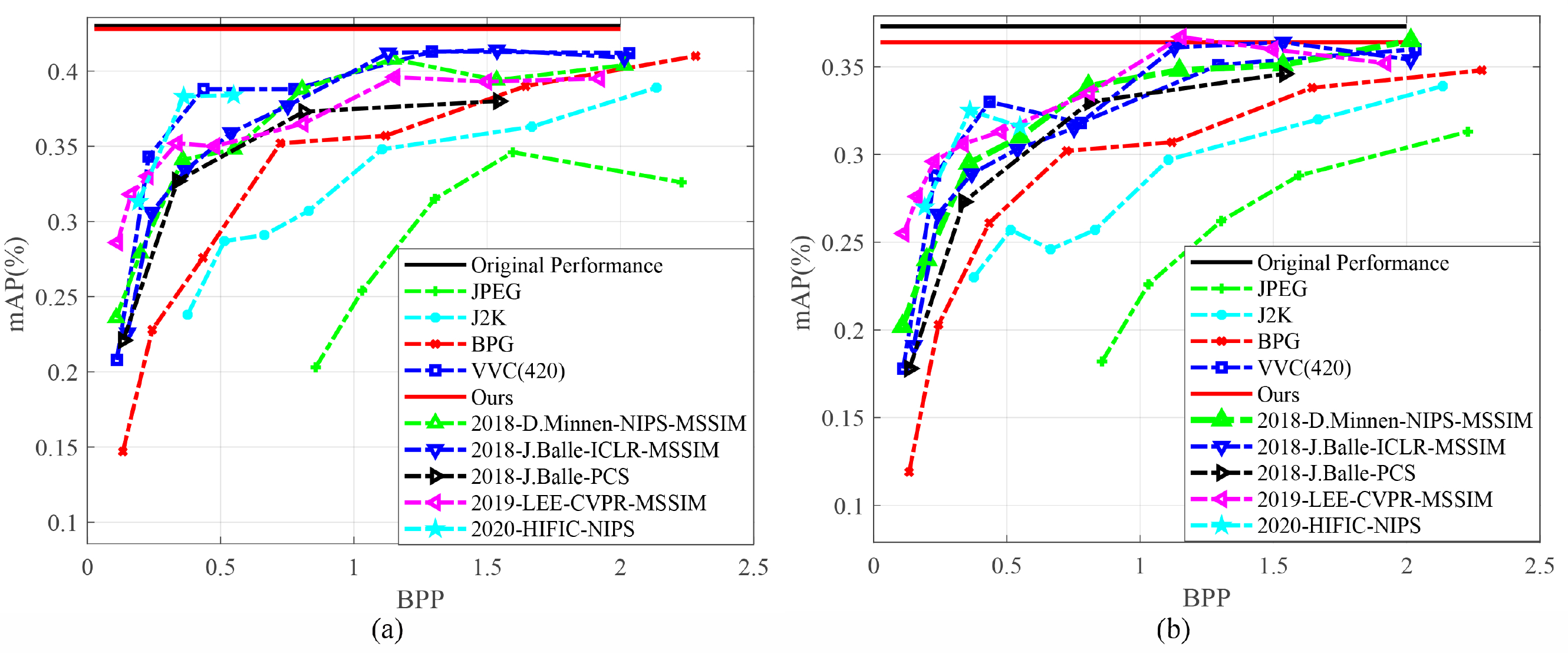}
		\caption{Performance of (a) object detection and (b) instance segmentation on the COCO2017.}\label{fig_map}
	\end{center}
\end{figure*}
\subsection{Datasets and evaluation metrics}
\label{Exp}
\subsubsection{Datasets}
Two datasets are involved in our experiments, i.e., COCO2017\cite{COCO2014} and Kodak\cite{kodak}. The COCO2017 dataset contains $80$ object categories, $118288$ Train images, $40670$ Test images, and $5000$ Val images. The Kodak dataset contains $24$ uncompressed PNG true-color images with a fixed resolution of $768\times512$. In this work, $8440$ high-resolution images (no less than $512$ pixels for each image) are selected from COCO2017 Train set to train the proposed model. For the testing phase, the first $50$ images from COCO2017 Val set and the entire $24$ images of Kodak are evaluated. 

\subsubsection{Evaluation metrics}
To evaluate the performance of the proposed codec for human consumption, we select PSNR and MS-SSIM \cite{wang2003multi} to measure the quality of the reconstructed image. Besides, we also use the mean Average Precision (mAP) to evaluate the performance of machine vision tasks, such as object detection and instance segmentation. It should be noticed that the mAP that we use here is the average AP value of $10$ Intersection over Union (IOU) thresholds, with an interval of 0.05 from 0.50 to 0.95. Commonly, It can better reflect the overall performance of the model compared with the mAP calculated by a single IOU.

\subsection{Anchors and implementation details}
\subsubsection{Anchors}
We compare our codec with $9$ typical codecs, which include $4$ traditional hybrid codecs (i.e., JPEG\cite{wallace1992the}, JPEG2000\cite{taubman2002jpeg2000}, BPG\cite{bellard2015bpg}, and VVC(420)\cite{vvc}) and $5$ deep-learnt codecs (i.e., 2018-D.Minnen-NIPS\cite{minnen2018joint}, 2018-J.Balle-ICLR\cite{balle2018variational}, 2018-J.Balle-PCS\cite{balle2018efficient}, 2019-Lee-CVPR\cite{lee2017context}, 2019-DSSLIC\cite{akbari2019dsslic}, 2020-HIFIC-NIPS\cite{mentzer2020high}. Since the DSSLIC\cite{akbari2019dsslic} does not provide the semantic segmentation map for the COCO2017 dataset, we only evaluate it on the Kodak dataset.

\subsubsection{Implementation details}
Before training, we first pre-process the input images by resizing them to $256\times256$ to guarantee that the input images have a fixed resolution for the subsequent training. Such pre-processing stage is saved for testing phase. In addition, the lossy codec \emph{VVC} and the lossless codec \emph{FLIF} are involved in the proposed paradigm. The default settings of \emph{VVC} (standardized test software VTM $8.2$) and \emph{FLIF} are used for testing. Besides, $6$ commonly used QPs ($17$, $22$, $27$, $32$, $37$, $42$) for \emph{VVC} are selected during our test. The total training epochs of the proposed model are set to $100$. For the first $50$ epochs, the learning rate is set to $0.02$. For the following $50$ epochs, the learning rate is gradually reduced. The batchsize is set to $8$. ADAM \cite{kingma2014adam} is used to optimize the model.


\begin{figure*}[htbp]
	\begin{center}
		\noindent
		\includegraphics[width = 7.2in]{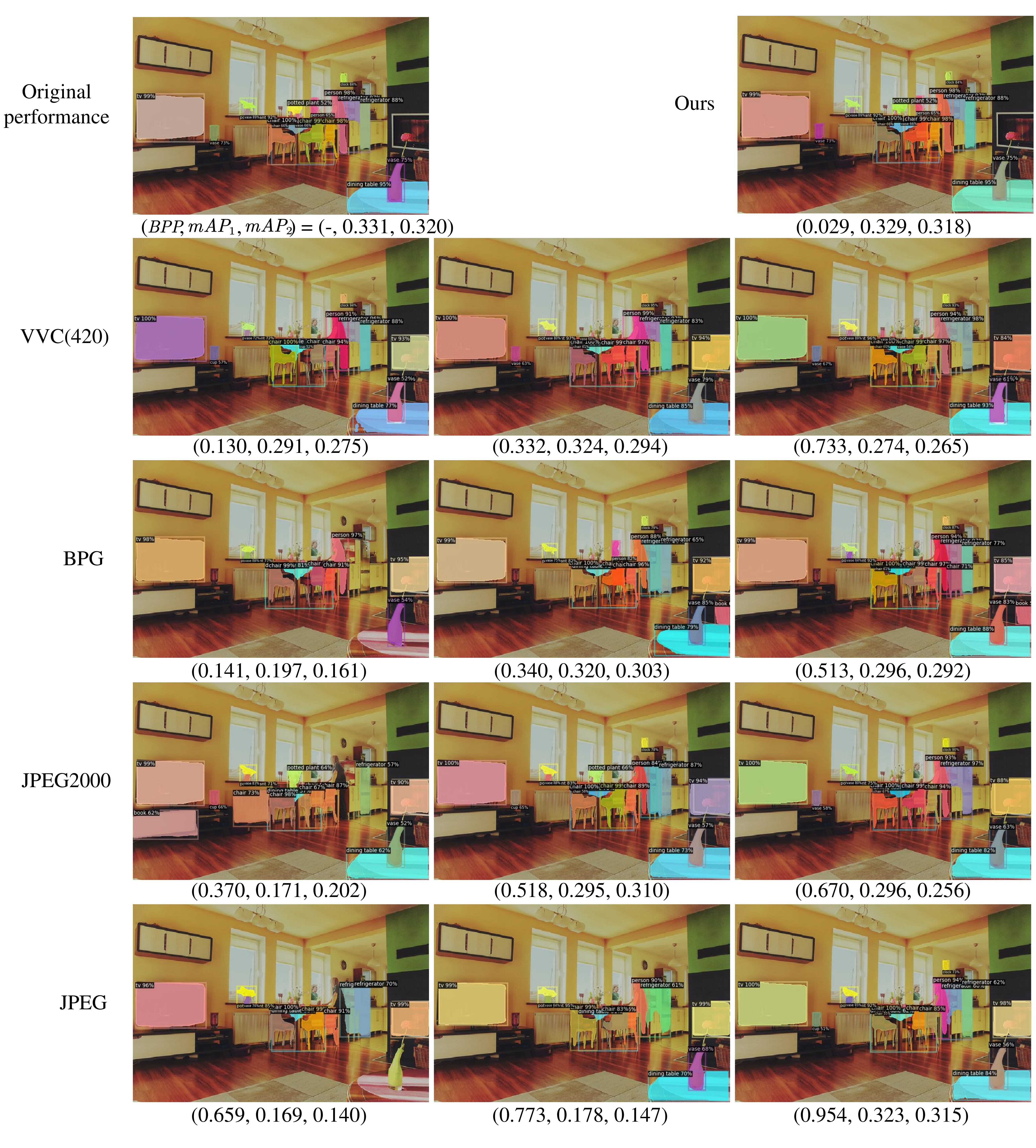}
		\caption{Visual comparison results of object detection and instance segmentation tasks.}\label{Fig. 11.}
\end{center}
\end{figure*}

\begin{figure*}[htbp]
	\begin{center}
		\noindent
		\includegraphics[width = 7.1in]{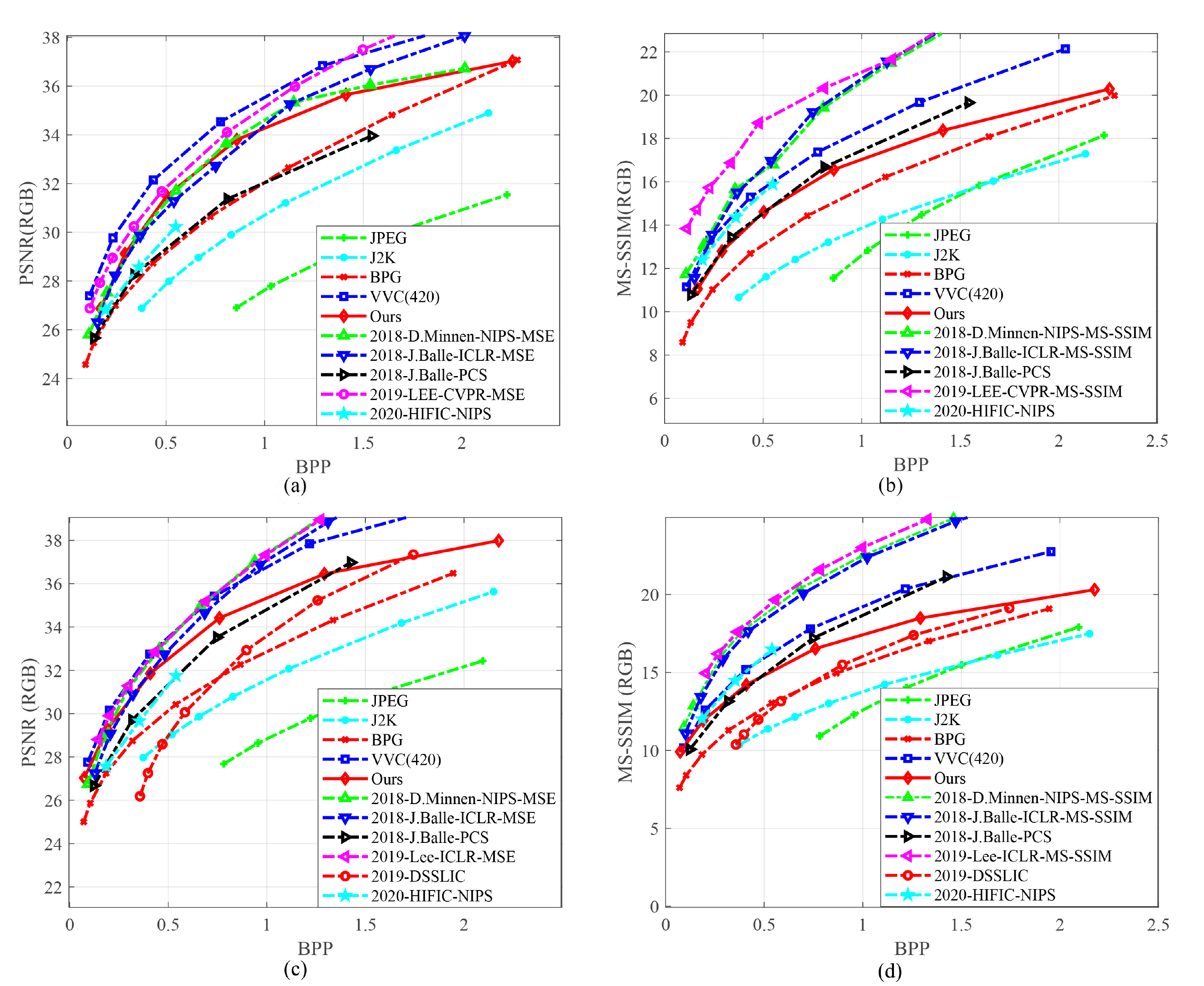}
		\caption{Performance of image reconstruction on two datasets. (a)-(b) and (c)-(d) are the test results on the COCO2017 and Kodak, respectively.}\label{Fig. 8.}
	\end{center}
\end{figure*}



\subsection{Evaluation on machine vision tasks}
For machine vision tasks, we only need to decoded the 16-bit gray-scale profile $\tilde{s}$ from $stream1$, since $\tilde{s}$ contains all the information for the object classification, object detection, and instance segmentation tasks. Besides, as $\tilde{s}$ is compressed with the lossless codec \emph{FLIF}, all the information above is well-preserved. Therefore, for the object classification task, it can be achieved by using the fully recovered category of object from $\tilde{s}$ without any performance degradation. 

For the object detection and instance segmentation tasks, we firstly compress the first $50$ image of COCO2017 Val set with the the proposed method and the anchors. Then, the compressed images together with their associated test images are fed into the Mask-RCNN with pre-trained weights to perform object detection and instance segmentation tasks. Fig. \ref{fig_map} exhibits the performance of our method and anchors on object detection and instance segmentation tasks in terms of bitrate vs. mAP. The BPP is computed by dividing total number of coded bits ($stream1$) by the number of input pixels. In the Fig. \ref{fig_map} (a) and (b), the black lines show the original mAP performance ($43\%$ for object detection and $37.3\%$ for instance segmentation) when using the test images as input to Mask-RCNN with pre-trained weights. The red lines represent our performance ($42.8\%$ for object detection and $36.4\%$ for instance segmentation), which are slightly lower than the original performance. Such insignificant performance degradation is mainly caused by rounding the floating coordinates of instance segmentation map into integers during 16-bit gray-scale representation. However, the performance of the proposed method outperforms that of the anchors, especially for the situations with low BPP. In other words, our method can still maintain high performance for machine vision tasks even for the low bitrate situations. It provides a feasible technique for machine-to-machine communications when bandwidth is limited.  

In addition, we also find that the mAP of some codecs don't increase with the increase of BPP. E.g., after BPP reaches $1.2$, the mAP of VVC starts to decrease. Since the VVC is designed for human consumption, it only takes human visual perceptual into account and regardless the characteristics of machine vision. Machine vision tasks are more sensitive to the semantic information degradation such as the position changes of the objects, and humans may more sensitive to the texture information degradation. This also demonstrate that the codecs designed for human consumption is not optimal for machine consumption. 

\begin{figure*}[htbp]
	\begin{center}
		\noindent
		\includegraphics[width = 7.0in]{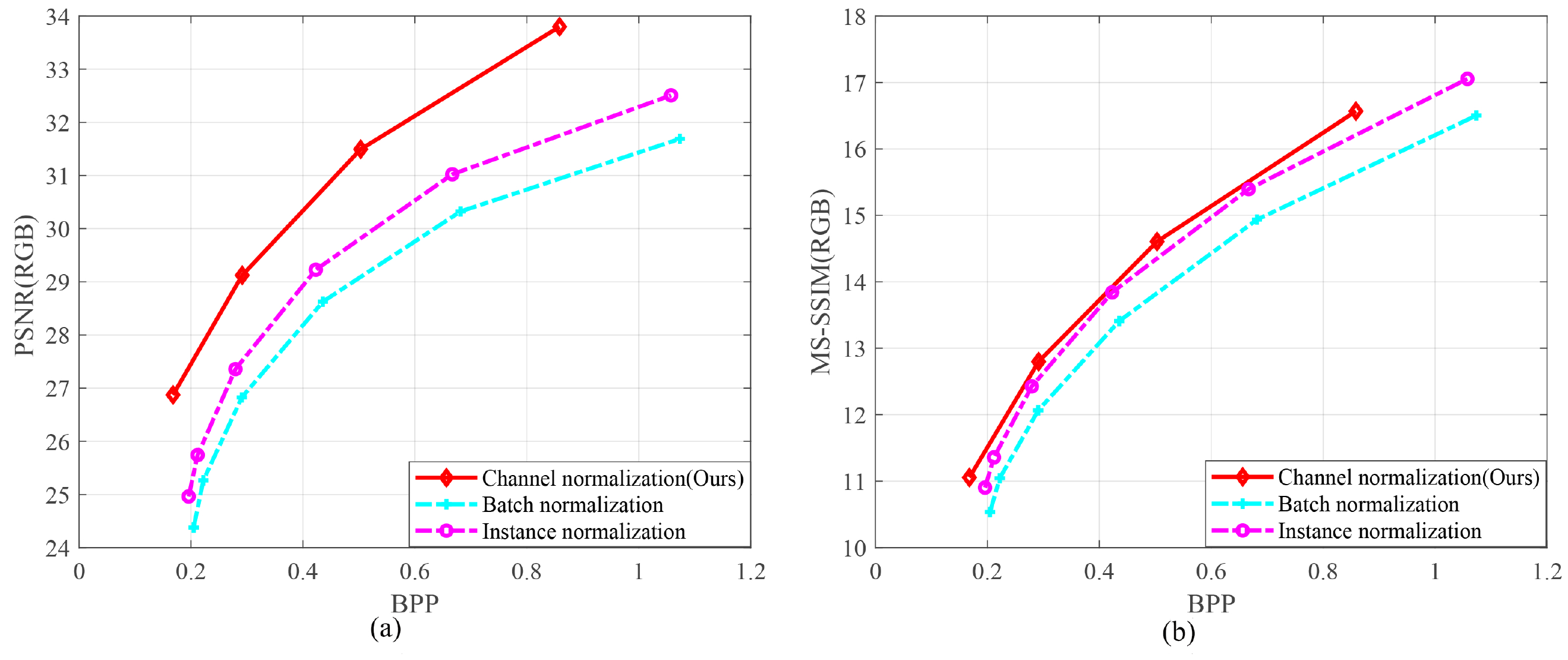}
		\caption{Performance of image reconstruction with different normalization methods on the COCO2017.}\label{fig_nor}
	\end{center}
\end{figure*}
\begin{figure}[htbp]
	\begin{center}
		\noindent
		\includegraphics[width = 3.4in]{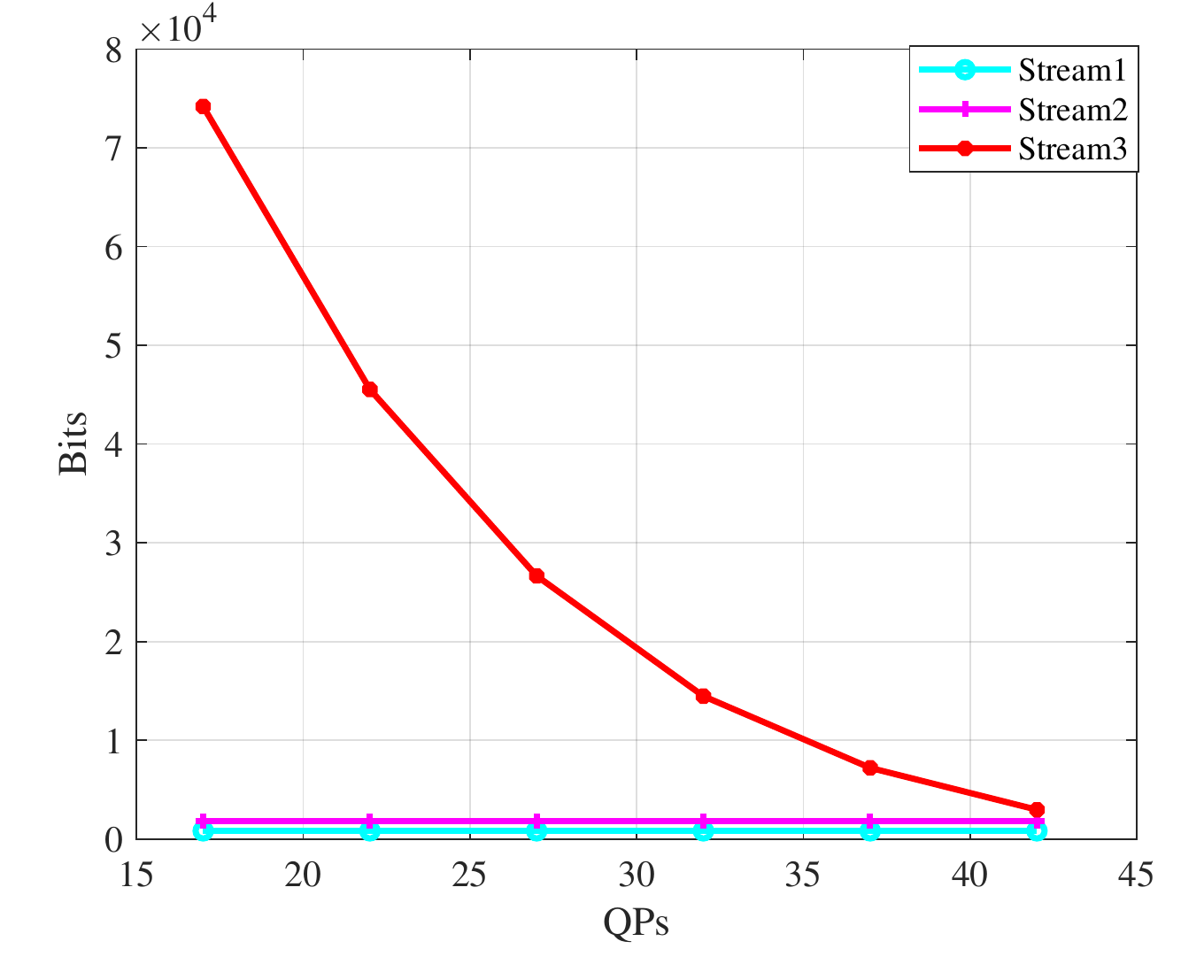}
		\caption{Bits cost of three different bitstreams under different QPs.}\label{fig_bit}
\end{center}
\end{figure}
The object detection and instance segmentation results of the proposed method and partial anchors (e.g., VVC, BPG, JPEG2000, JPEG) are visualized in Fig. \ref{Fig. 11.}. For each codec, we select three different BPPs and visualize their corresponding results on object detection and instance segmentation. Based on the observation of this figure, we can demonstrate that the proposed method achieves the closest result to the original performance. Besides, the BPP, mAP for object detection and instance segmentation are also listed in ($BBP$, $mAP_1$, $mAP_2$).


\subsection{Evaluation on image reconstruction}
To evaluate the performance of image reconstruction, the proposed method and the anchors are examined on the first $50$ images of COCO2017 Val set and the entire $24$ images of Kodak. For the proposed method, the entire bitstreams (including $stream1$, $stream2$, and $stream3$) are decoded for image reconstruction.  

The performance of image reconstruction are shown in Fig. \ref{Fig. 8.}. Similarly, the BPP is computed by dividing total number of coded bits (three bitstreams) by the number of input pixels. Fig. \ref{Fig. 8.} (a)-(b) and (c)-(d) show the performance of image reconstruction on COCO2017 dataset and Kodak dataset. In terms of PSNR, the proposed method outperforms most traditional hybrid codecs (such as JPEG, JEPG2000, BPG) and deep-learnt codecs (2018-J.Balle-ICLR, 2018-D.Minnen-NIPS, Mentzer2020, Balle2018-PCS, Balle2018-PCS, DSSLIC), especially for the low BPP situations. In terms of MS-SSIM, the proposed method still achieves comparable results of most codecs. 

\begin{figure*}[htbp]
	\begin{center}
		\noindent
		\includegraphics[width = 7.0in]{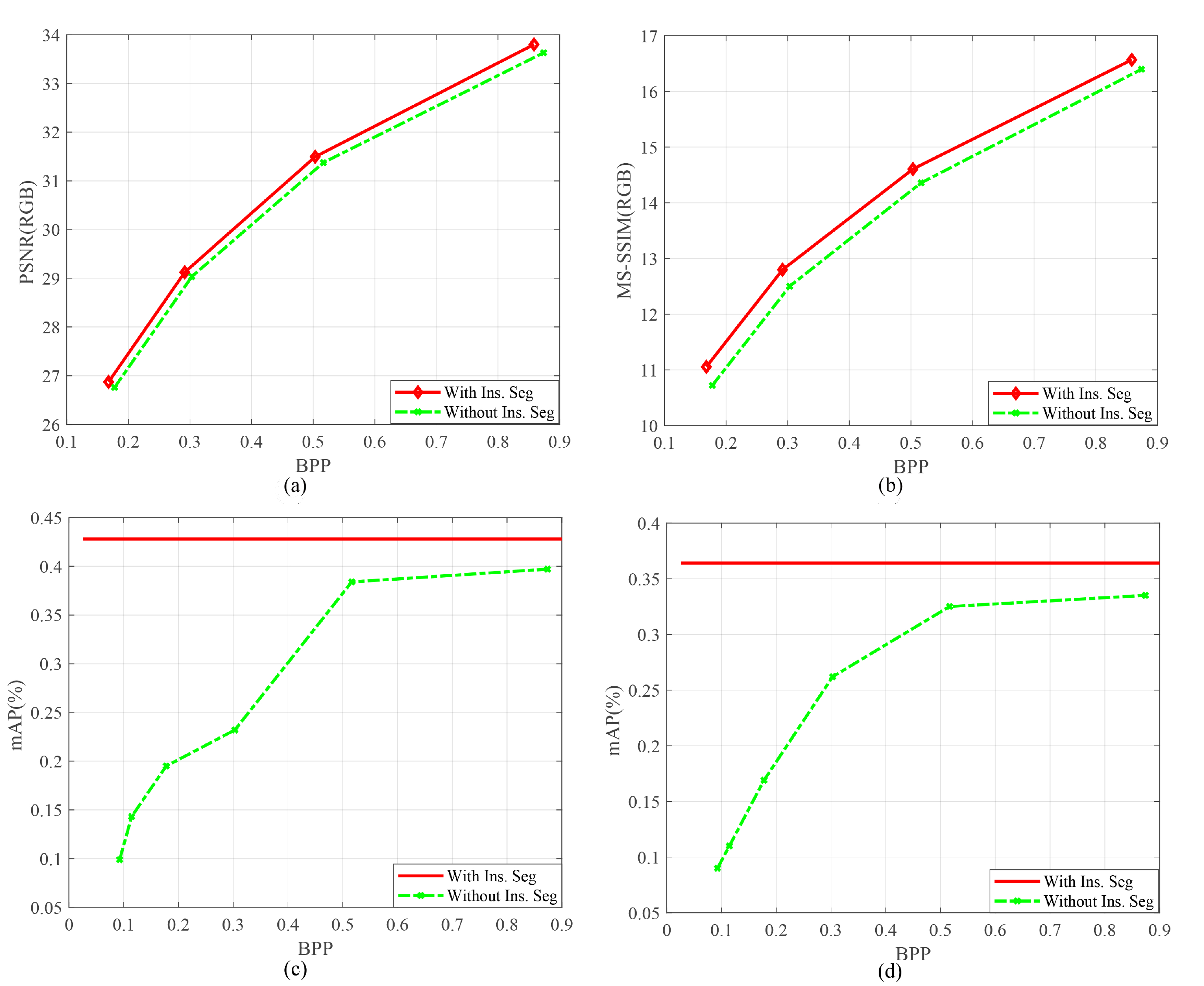}
		\caption{Performance of image reconstruction and machine vision tasks with/without the help of high-level instance segmentation map. (a) and (b) are the image reconstruction results when the proposed method is performed with/without the high-level instance segmentation map. Similarly, (c) and (d) are the object detection and instance segmentation results.}\label{fig_ab}
\end{center}
\end{figure*}
\subsection{Evaluation on different normalization methods}
\label{nor}
Normalization is one of the critical components in the proposed module \emph{LE} and \emph{IP}, which plays a key role in improving the compression efficiency for image reconstruction. In this subsection, we test three typical normalization methods (such as channel normalization, batch normalization, and instance normalization). This can be achieved by replacing the $Nor$ in Fig. \ref{fig_LE} and \ref{fig_IP} with channel normalization, batch normalization, and instance normalization, respectively. The results are shown in Fig. \ref{fig_nor}. For the image reconstruction, channel normalization significantly improves the compression efficiency of the proposed method compared with the other two normalization methods in terms of PSNR and MS-SSIM. These results demonstrate our speculates in subsection \ref{LE_IP}, i.e, much more redundancy along channel dimension (maybe include some useful information for image reconstruction) is removed at the \emph{LE} and the over-compressed information may be recovered at the \emph{IP} during image reconstruction. Channel normalization will benefit the process above. Therefore, it's more suitable for the proposed predictive coding paradigm.

\subsection{Bitstream analysis}
Fig. \ref{fig_bit} shows the bits cost of three different bitstreams under different QPs. Since the first two bitstreams ($stream1$ and $stream2$) are obtained with lossless codec FLIF, their bits cost are fixed. Besides, the bits cost of $stream1$ is a little lower than that of $stream2$. The third bitstream $stream3$ is obtained with the lossy codec VVC. Hence, it has high/low bits cost when compressed with small/large QPs. 

\subsection{Ablation experiment}
To verify that the the high-level instance segmentation map representation $\tilde{s}$ plays role in both image reconstruction and machine vision tasks, we design an ablation experiment in this subsection. Firstly, we remove the \emph{HE} and \emph{RE} from the paradigm. Correspondingly, the $\tilde{s}$ is also removed from the inputs of \emph{LE} and \emph{IP}. After that, the \emph{LE} and \emph{IP} are retrained. Then, we compress the low-level signal features $y$ and residual map $r$ with lossless codec \emph{FLIF} and lossy codec $VVC$, respectively. At the decoder side, they are decoded to $y$ and $\tilde{r}$, respectively. Then, both $y$ and $\tilde{r}$ are used to perform image reconstruction. The reconstructed image are further used to perform object detection and instance segmentation. 

The performance of image reconstruction with/without instance segmentation map are shown in Fig. \ref{fig_ab} (a) and (b). The results demonstrate that high-level instance segmentation map information $\tilde{s}$ improves the quality of the reconstructed image in terms of PSNR and MS-SSIM. Besides, the performance of object detection and instance segmentation with/without instance segmentation map are shown in Fig. \ref{fig_ab} (c) and (d). The results also demonstrate that high-level instance segmentation map information $\tilde{s}$ improves the accuracy of object detection and instance segmentation in terms of mAP. All these results suggest that the high-level instance segmentation map information $\tilde{s}$ improves the performance of both image reconstruction and machine vision tasks. In other words, it improves the compression efficiency of the proposed image codec for both humans and machines.

\section{Conclusion}

In this work, we have proposed a feasible image codec paradigm for both human and machine uses. Unlike extracting and representing the high-level semantic features for performing machine vision tasks. The instance segmentation map is extracted and represented with a more compact 16-bit gray-scale profile. With this profile, on the one hand, we can directly achieve high-performance machine vision tasks (e.g., object classification, object detection, and instance segmentation); on the other hand, it helps to predict/reconstruct a general-quality image together with the extracted low-level signal features via our proposed image predictor. Besides, the residual information between the original image and its associated predicted one is compressed and used for high-quality image reconstruction. All these designs enable us to achieve scalable image compression. Experimental results show that our method achieves comparable performance as most learning-based codecs and outperforms the traditional codecs (e.g., BPG and JPEG2000) for image reconstruction while outperforming the existing codecs for various machine vision tasks.

\bibliographystyle{IEEEtran}

\begin{thebibliography}{}
\providecommand{\url}[1]{#1}
\csname url@samestyle\endcsname
\providecommand{\newblock}{\relax}
\providecommand{\bibinfo}[2]{#2}
\providecommand{\BIBentrySTDinterwordspacing}{\spaceskip=0pt\relax}
\providecommand{\BIBentryALTinterwordstretchfactor}{4}
\providecommand{\BIBentryALTinterwordspacing}{\spaceskip=\fontdimen2\font plus
\BIBentryALTinterwordstretchfactor\fontdimen3\font minus
  \fontdimen4\font\relax}
\providecommand{\BIBforeignlanguage}[2]{{%
\expandafter\ifx\csname l@#1\endcsname\relax
\typeout{** WARNING: IEEEtran.bst: No hyphenation pattern has been}%
\typeout{** loaded for the language `#1'. Using the pattern for}%
\typeout{** the default language instead.}%
\else
\language=\csname l@#1\endcsname
\fi
#2}}
\providecommand{\BIBdecl}{\relax}
\BIBdecl

\end{thebibliography}


\begin{thebibliography}{99}
\bibitem{sneyers2016flif}
J. Sneyers and P. Wuille, “FLIF: Free lossless image format based on maniac compression,” in \emph{2016 IEEE International Conference on Image Processing (ICIP)}. IEEE, pp. 66–70, 2016.

\bibitem{wang2003multi}
Z. Wang, E. P. Simoncelli, and A. C. Bovik, “Multiscale structural similarity for image quality assessment,” in \emph{Proc. IEEE Conf. Rec. 37th Asilomar Conf. Signals, Syst. Comput}., vol. 2, pp. 1398–
1402, 2003.

\bibitem{kodak}
Kodak PhotoCD dataset. [Online]. Available: http://r0k.us/graphics/kodak/

\bibitem{COCO2014}
T.-Y. Lin, M. Maire, S. Belongie, J. Hays, P. Perona, D. Ramanan,P. Dollar, and C. L. Zitnick. Microsoft COCO: Common objects in context. In \emph{ECCV}. 2014.

\bibitem{vvc}
H. Fraunhofer, “Vvc official test model vtm.” [Online]. Available: https: //vcgit.hhi.fraunhofer.de/jvet/VVCSoftware VTM/tree/VTM-8.2

\bibitem{kirillov2019detectron2}
Yuxin Wu, Alexander Kirillov, Francisco Massa, Wan-Yen Lo, and Ross Girshick. Detectron2.[Online]. Available: https://github.com/facebookresearch/detectron2, 2019.

\bibitem{duan2020video}
L. Duan, J. Liu, W. Yang, T. Huang, and W. Gao, “Video coding for machines: A paradigm of collaborative compression and intelligent analytics,”\emph{IEEE Trans. Image Process}., vol. 29, pp. 8680–8695, 2020.

\bibitem{lowe2007distinctive}
D. Lowe, “Distinctive Image Features from Scale-Invariant Keypoints,” Int’l J. \emph{Computer Vision}, vol. 2, no. 60, pp. 91-110, 2004.

\bibitem{Djamasbi2004compact}
 L.-Y. Duan, J. Lin, J. Chen, T. Huang, and W. Gao, “Compact descriptors for visual search,” \emph{IEEE Multimedia}, vol. 21, no. 3, pp. 30–40,
Jul./Sep. 2014.

\bibitem{BO2019direct}
I. K. S. Bo Shen, “Direct feature extraction from compressed images,” Storage Retr. Still Image Video Databases IV, vol. 2670, pp. 404–414, 2019.

\bibitem{qian2012an}
Z. Qian, W. T. Qiao, An Edge Detection Method in DCT Domain. \emph{Procedia Engineering}, vol. 29, pp. 344-348, 2011.

\bibitem{khatoonabadi2013video}
S. Khatoonabadi and I. Bajic, “Video object tracking in the compressed domain using spatio-temporal Markov random fields,” \emph{IEEE Trans. Image Process}., vol. 22, no. 1, pp. 300–313, Jan. 2013

\bibitem{choi2017corner}
H. Choi and I. V. Bajic, “Corner proposals from HEVC bitstreams,” in Proc. \emph{IEEE ISCAS}, pp. 1–4, 2017.

\bibitem{choi2017hevc}
H. Choi and I. V. Bajic, “HEVC intra features for human detection,” in Proc. \emph{IEEE GlobalSIP}, pp. 393–397, 2017.

\bibitem{ranjbar2018can}
S. R. Alvar, H. Choi, and I. V. Bajic, “Can you tell a face from a HEVC bitstream?,” in \emph{Proc. IEEE MIPR’18}, Apr. 2018

\bibitem{toderici2016variable}
G. Toderici et al. “Variable rate image compression with recurrent neural networks.” [Online]. Available: https://arxiv.org/abs/1511.06085, 2015.

\bibitem{balle2017end}
J. Ballé, V. Laparra, and E. P. Simoncelli. “End-to-end optimized image compression.” [Online]. Available: https://arxiv.org/abs/1611.01704, 2016.

\bibitem{toderici2017full}
G. Toderici et al. “Full resolution image compression with recurrent neural networks.” [Online]. Available: https://arxiv.org/abs/1608.05148, 2016.

\bibitem{johnston2018improved}
 N. Johnston et al. “Improved lossy image compression with priming and spatially adaptive bit rates for recurrent networks.” [Online].Available: https://arxiv.org/abs/1703.10114, 2017.

\bibitem{theis2017lossy}
L. Theis, W. Shi, A. Cunningham, and F. Huszár. “Lossy image compression with compressive autoencoders.” [Online]. Available: https://arxiv.org/abs/1703.00395, 2017.

\bibitem{agustsson2017soft}
E. Agustsson et al., “Soft-to-hard vector quantization for end-toend learning compressible representations,” in \emph{Proc. Adv. Neural Inf. Process. Syst}., pp. 1141–1151, 2017.


\bibitem{balle2018variational}
J. Ballé, D. Minnen, S. Singh, S. J. Hwang, and N. Johnston, “Variational image compression with a scale hyperprior,” in\emph{ Proc. Int. Conf. Learn. Represent}., pp. 1–10, May 2018.

\bibitem{minnen2018joint}
D. Minnen, J. Ballé, and G. D. Toderici, “Joint autoregressive and hierarchical priors for learned image compression,” in \emph{Proc. Adv. Neural Inf. Process. Syst}., pp. 10771–10780, 2018.

\bibitem{cheng2019Deep}
Z. Cheng, H. Sun, M. Takeuchi, and J. Katto, “Deep Residual Learning for Image Compression,”  [Online]. Available: http://arxiv.org/abs/1906.09731, 2019,.

\bibitem{bellard2015bpg}
F. Bellard. The BPG Image Format, [Online]. Available: http://bellard.org/bpg/, accessed on Sep. 20, 2015.

\bibitem{wallace1992the}
G. K. Wallace, “The JPEG still picture compression standard,” \emph{Commun. ACM}, vol. 34, no. 4, pp. 30–44, 1991.

\bibitem{taubman2002jpeg2000}
D. S. Taubman and M. W. Marcellin, JPEG2000: Image Compression Fundamentals, Standards, and Practice. Norwell, MA: Kluwer, 2001.

\bibitem{torfason2018towards}
R. Torfason, F. Mentzer, E. Agustsson, M. Tschannen, R. Timofte, and L. Van Gool. “Towards image understanding from deep compression without decoding.” [Online]. Available: https://arxiv.org/abs/1803.06131, 2018.

\bibitem{alvar2019multi}
S. R. Alvar and I. V. Bajic, “Multi-task learning with compressible features for collaborative intelligence,” in \emph{Proc. IEEE Int. Conf. Image Process. (ICIP)}, Sep.pp. 1705–1709, 2019.

\bibitem{patwa2020semantic}
N. Patwa, N. Ahuja, S. Somayazulu, O. Tickoo, S. Varadarajan, and S. Koolagudi, “Semantic-Preserving Image Compression,” \emph{Proc. - Int. Conf. Image Process. ICIP}, vol. 2020-October, pp. 1281–1285, 2020.

\bibitem{alvar2020bit}
S. R. Alvar and I. V. Baji, “Bit allocation for multi-task collaborative intelligence,” in \emph{Proc. IEEE Conf. Acoust., Speech, Signal Process}., pp. 4342–4346, 2020.

\bibitem{hu2020towards}
Y. Hu, S. Yang, W. Yang, L.-Y. Duan, and J. Liu, “Towards
coding for human and machine vision: A scalable image coding approach,” Jan. [Online]. Available:http://arxiv.org/abs/2001.02915, 2020,

\bibitem{liu2021semantics}
K. Liu, D. Liu, L. Li, N. Yan, and H. Li, “Semantics-to-signal scalable image compression with learned revertible representations,” \emph{Int. J. Comput. Vis}., pp. 1–17, Jun. 2021.

\bibitem{choi2021latent}
H. Choi and I. V. Bajic, “Latent-space scalability for multi-task collaborative intelligence,” in \emph{IEEE ICIP}, arXiv:2105.10089, 2021.

\bibitem{ren2017faster}
S. Ren, K. He, R. Girshick, and J. Sun, “Faster R-CNN: Towards Real-Time Object Detection with Region Proposal Networks,” IEEE Trans. Pattern Anal. Mach. Intell., vol. 39, no. 6, pp. 1137–1149, 2017.

\bibitem{zhao2017pyramid}
H. Zhao, J. Shi, X. Qi, X. Wang, and J. Jia, “Pyramid scene parsing network,” in \emph{Proc. IEEE Int. Conf. Comput. Vis}., Jul. pp. 2881–2890, 2017.

\bibitem{he2017mask}
K. He et al., “Mask R-CNN,” in \emph{Proc. ICCV}, pp. 2980–2988, 2017.

\bibitem{cheng2020learned}
 Z. Cheng, H. Sun, M. Takeuchi, and J. Katto, “Learned image compression with discretized Gaussian mixture likelihoods and attention modules,” [Online]. Available: http://arxiv.org/abs/2001.01568, 2020.

\bibitem{luo2018deepsic}
S. Luo, Y. Yang, Y. Yin, C. Shen, Y. Zhao, and M. Song, “DeepSIC:Deep semantic image compression,” in \emph{Proc. ICONIP 2018: Neural Inf.Process}., pp. 96–106, 2018.

\bibitem{akbari2019dsslic}
M. Akbari, J. Liang, and J. Han, “DSSLIC: Deep semantic
segmentation-based layered image compression,” in \emph{Proc. IEEE Int.Conf. Acoust., Speech Signal Process}., pp. 2042–2046, 2019.

\bibitem{agustsson2019generative}
E. Agustsson, M. Tschannen, F. Mentzer, R. Timofte, and L. Van Gool, “Generative adversarial networks for extreme learned image compression,”  [Online]. Available: http://arxiv.org/abs/1804.02958, 2018.

\bibitem{mentzer2020high}
F. Mentzer, G. D. Toderici, M. Tschannen, and E. Agustsson, “Highfidelity generative image compression,” in \emph{Proc. Adv. Neural Inf. Process. Syst}., pp. 11913–11924, 2020.

\bibitem{wang2004image}
Z. Wang, A. C. Bovik, H. R. Sheikh, and E. P. Simoncelli,
“Image quality assessment: From error visibility to structural similarity,” \emph{IEEE Trans. Image Process}., vol. 13, no. 11, pp. 600–612, Apr. 2004.

\bibitem{balle2018efficient}
J. Ballé, “Efficient nonlinear transforms for lossy image compression,” in Proc. PCS., pp. 248–252, Jun. 2018.

\bibitem{ulyanov2016instance}
D. Ulyanov, A. Vedaldi, and V. Lempitsky. “Instance normalization: The missing ingredient for fast stylization.” [Online]. Available: https://arxiv.org/abs/1607.08022, 2016. 

\bibitem{lee2017context}
J. Lee, S. Cho, and S.-K. Beack, “Context-adaptive entropy model for end-to-end optimized image compression,” in \emph{Proc. 7th Int. Conf. Learn. Representation}, 2019.

\bibitem{JND}
J. Jin, “Just Noticeable Difference for Deep Machine Vision," in \emph{IEEE Transactions on Circuits and Systems for Video Technology}, doi: 10.1109/TCSVT.2021.3113572, 2021.

\bibitem{rippel2017real}
O. Rippel and L. Bourdev, “Real-time adaptive image compression,” in \emph{International Conference on Machine Learning}, pp. 2922–2930, 2017.

\bibitem{Santurkar2018g}
S. Santurkar, D. Budden, N. Shavit, “Generative Compression”, \emph{Picture Coding Symposium}, June 24-27, 2018.

\bibitem{li2018l}
M. Li, W. Zuo, S. Gu, D. Zhao, D. Zhang, “Learning Convolutional Networks for Content-weighted Image Compression”, \emph{IEEE Conf. on Computer Vision and Pattern Recognition (CVPR)}, June 17-22, 2018.

\bibitem{chneg2018d}
Z. Cheng, H. Sun, M. Takeuchi, J. Katto, “Deep Convolutional AutoEncoder-based Lossy Image Compression”, \emph{Picture Coding Symposium}, pp. 1-5, June 24-27, 2018.

\bibitem{mentzer2018c}
F. Mentzer, E. Agustsson, M. Tschannen, R. Timofte, L. V. Gool, “Conditional Probability Models for Deep Image Compression”, \emph{IEEE Conf. on Computer Vision and Pattern Recognition (CVPR)}, June 17-22, 2018.

\bibitem{choi2021scalable}
H. Choi and I. V. Bajic, “Scalable image coding for humans and machines,” 2021, arXiv:2107.08373. [Online]. Available: http://arxiv.org/abs/2107.08373


\bibitem{yan2021scalable}
N. Yan, C. Gao, D. Liu, H. Li, L. Li and F. Wu, SSSIC: Semantics-to-Signal Scalable Image Coding With Learned Structural Representations, \emph{IEEE Transactions on Image Processing}, vol. 30, pp. 8939-8954, 2021.

\bibitem{dodge2016u}
S. Dodge and L. Karam, Understanding how image quality affects deep neural networks, \emph{2016 Eighth International Conference on Quality of Multimedia Experience (QoMEX)}, pp. 1-6, 2016.

\bibitem{sun2021semantic}
S. Sun, T. He and Z. Chen, "Semantic Structured Image Coding Framework for Multiple Intelligent Applications," in \emph{IEEE Transactions on Circuits and Systems for Video Technology}, vol. 31, no. 9, pp. 3631-3642, Sept. 2021.

\bibitem{he2019beyond}
T. He, S. Sun, Z. Guo and Z. Chen, "Beyond Coding: Detection-driven Image Compression with Semantically Structured Bit-stream," \emph{2019 Picture Coding Symposium (PCS)}, pp. 1-5, 2019 .

\bibitem{hy2021revisit}
Hu Y, Yang W, Huang H, et al. Revisit Visual Representation in Analytics Taxonomy: A Compression Perspective[J]. \emph{arXiv preprint} arXiv:2106.08512, 2021.

\bibitem{sun2020cdva}
B. Sun, H. Sha, M. Rafie and L. Yang, "CDVA/VCM: Language for Intelligent and Autonomous Vehicles," \emph{2020 IEEE International Conference on Image Processing (ICIP)}, pp. 3104-3108, 2020.

\bibitem{kingma2014adam}
D.~P. Kingma and J.~Ba, ``Adam: A method for stochastic optimization,'' {\em
  arXiv preprint arXiv:1412.6980}, 2014.

\end{thebibliography}
%
%
%

%




\end{document}